\documentclass[12pt]{JHEP3}
\usepackage{amsfonts,amssymb}
\usepackage{amsmath}

\usepackage{epsf}

\def\T11{{T}^{1,1}}
\def\bear{\begin{eqnarray}}
\def\eear{\end{eqnarray}}

\newcommand{\vac}{{|0\rangle}}

\def\Vol{\mathrm{Vol}}

\newcommand{\pa}{\partial}
\newcommand{\tr}{{\rm tr}}
\newcommand{\comment}[1]{}

\newcommand{\pasl}{\pa\kern-.55em /}

\newcommand{\ksl}{k\kern-.55em /}

\newcommand{\ket}[1]{|#1\rangle}
\newcommand{\bra}[1]{\langle #1|}
\newcommand{\braket}[2]{\langle #1|#2\rangle}

\DeclareFixedFont{\xiiss}{OT1}{cmss}{m}{n}{12}
\DeclareFixedFont{\ixss}{OT1}{cmss}{m}{n}{9}
\DeclareFixedFont{\cmrnine}{OT1}{cmr}{m}{n}{9}
\newcommand{\field}[1]{\mathbb{#1}}

\newcommand{\BR}{{\field R}}

\newcommand{\CCs}{\hbox{\ixss C\kern-.4emI}}
\newcommand{\ZZs}{\hbox{\ixss Z\kern-.4emZ}}

\newcommand{\myfig}[3]{\begin{figure}[ht]
\begin{center}
\leavevmode
\epsfxsize=#2cm
\epsfbox{#1}
\end{center}
\caption{#3}
\label{fig:#1}
\end{figure}}

\title{All loop BMN state energies from matrices}
\author{David Berenstein$^{\dagger,1}$, Diego H. Correa$^{\ddagger,2}$,
Samuel E. V\'azquez$^{\dagger,3}$\\
$^\dagger$ Department of Physics, UCSB, Santa Barbara, CA 93106\\
$^\ddagger$ Kavli Institute for theoretical Physics, Santa Barbara, CA 93106
\\
$^1$ \email{dberens@physics.ucsb.edu}
\\
$^2$ \email{dcorrea@kitp.ucsb.edu}
\\
$^3$ \email{svazquez@physics.ucsb.edu}}

\abstract{We study a quantum corrected 
$SO(6)$ invariant  matrix quantum mechanics obtained
from the $s$-wave modes of the scalars of ${\cal N} = 4$ SYM on
$S^3$. For commuting matrices, this model is believed to describe
the 1/8 BPS states of the full SYM theory. In the large $N$ limit the
ground state corresponds to a distribution of eigenvalues on a
$S^5$ which we identify with the sphere on the dual geometry
$AdS_5\times S^5$. We then consider non-BPS excitations by studying
 matrix perturbations where the off-diagonal modes are treated perturbatively.
 To a first approximation, these modes can be described by a free theory of
``string bits"  whose energies depend on the diagonal degrees of freedom.
 We then consider a state with two string bits and large 
angular momentum $J$ on the sphere. In the large $J$ limit
we use a simple saddle point approximation to show that the energy
of these states coincides precisely with the BMN spectrum to all orders in the
't Hooft coupling. We also find some new problems with the all loop Bethe Ansatz 
conjecture of the ${\cal N}=4$ SYM planar spin chain model.}

\preprint{NSF-KITP-05-82}

\begin{document}

\section{Introduction}

The correspondence between IIB string theory on the background $AdS_5\times S^5$
and the conformal ${\cal N}=4$ super Yang-Mills theory \cite{Malda}, is the best understood
realization of the  effective string theory description of large $N$ strongly coupled gauge theories proposed by 't Hooft in \cite{thooft}. Indeed, in \cite{Malda} this effective description was upgraded to an exact duality between the string theory description and the field theory
dynamics. 
One of the intriguing implications of this duality is that classical gravitational
physics of  $AdS_5\times S^5$ should be somehow encoded in the strong coupling
regime of the dual gauge field theory.

In a recent work \cite{Droplet}, a qualitative description of how the $AdS_5\times S^5$
geometry and locality appear in the strong 't Hooft coupling regime and large $N$ limit
of ${\cal N}=4$ SYM was presented, at least as far as the physics of the five sphere 
at the origin of $AdS$ is concerned . This derivation was carried by studying 1/8 BPS
configurations by means of a quantum matrix model  obtained from the $s$-wave modes
of the ${\cal N}=4$ SYM scalar fields when compactified on a round $S^3$. This model 
needs to be quantum corrected by hand because it is not supersymmetric, but if one assumes the result that would be obtained by using supersymmetry, in the end 
 this turns into a matrix model for 3 commuting complex matrices,
whose dynamics is described in terms of their eigenvalues. We will discuss this 
correction in more detail in the bulk of the paper.
These eigenvalues can be
effectively accounted as $N$ bosons in a six dimensional phase space with repulsive
interactions, where the phase space symplectic form is induced from studying dynamical solutions which respect the given amount of supersymmetry. Considering these bosons as a statistical ensemble and performing a saddle
point approximation, it is possible to obtain a density distribution of the eigenvalues.
Then, in the ground state, bosons turn out to be uniformly distributed
in a 5-sphere of the 6 dimensional phase space. This 5-sphere can be directly identified
with the 5-sphere of the $AdS_5\times S^5$ geometry. It is also possible to consider non-BPS configurations in this setup by turning on
off-diagonal modes for the matrices. An off-diagonal mode should involve two eigenvalues
and we can depict it with a straight arrow between the eigenvalues in the droplet. We call these {\em string bits}.  This way of distinguishing eigenvalues having some extra energy can be interpreted as a way of
localizing massive string bits in the geometrical $S^5$.

In this paper we provide quantitative evidence confirming that the picture of the
previous paragraph is correct not just at the qualitative level, but that it can be used to reproduce some results known by other means. The main
result we present is the calculation of the energy of BMN states \cite{BMN} to all 
orders of perturbation theory, summing only the planar  diagrams.  Our calculations should be
considered as  the strong coupling regime of the field theory. This result has been
originally reported in \cite{SZ} by using a very different argument.
 However, some steps on that derivation are not completely
justified, such as the use of the equations of motion of SYM without a complete analysis of contact terms that could spoil the relations used. In \cite{BDS}, assuming integrability
of the dilatation operator to all perturbative orders, a similar result which was more general was proposed. In this
case the result included two arbitrary constants that had to be fixed. Our results are extremely efficient at producing the energies of these states, especially when compared to a two loop calculation, as in \cite{GMR} and we will see that we are able to reproduce some of the conjectures made in \cite{BDS}. Also, one can generalize our results very easily to include multi-impurity states in the BMN limit also reproducing the square root formulas exactly without much effort.

The literature on this subject is vast and a complete review of all results that are related to the subject is 
prohibitive.  A good discussion and a comprehensive collection of references can be found in 
 Niklas Beisert's doctoral thesis \cite{Bphysrept}. A more introductory set of notes can be found in \cite{general}. We will refer the reader to these works for a complete
 guide to the literature.

The paper is organized as follows. In section \ref{mqm} we review the matrix model of
commuting matrices describing BPS states in ${\cal  N}=4$ SYM and the emergence
of the 5-sphere where the eigenvalues are distributed. We also compute in this section
the exact radius for the sphere. In section \ref{spa} we use a saddle point
approximation to compute the energy of BMN states to all orders in perturbation theory.
Finally, in section \ref{dis} we discuss our results, the validity of the approximations
we make and the possibility of extending similar computations beyond the BMN limit.

\section{Gauged matrix quantum mechanics of commuting matrices}
\label{mqm}

Let us consider a matrix quantum mechanics model for $2d$ Hermitian matrices  of rank
$N$ that commute with each other (we can equally consider it as a matrix model for $d$
normal  matrices that commute with each other). As argued in \cite{Droplet} such a
model results from considering either $1/2, 1/4$ or $1/8$ BPS states in ${\cal  N}=4$
SYM compactified on a sphere, where $d=1,2,3$ respectively. The BPS states near the vacuum
are made of multiple gravitational quanta, so they can be described in a purely geometric
fashion. In this section we will deal with the systems where the matrices commute without
describing how the other degrees of freedom of the ${\cal  N}=4$ SYM decouple. We 
will address this issue in the next section.

Let us label the Hermitian matrices by $X^a$, $a=1,\dots 2d$, and the complex normal
matrices by $Z^a= X^{2a-1}+i X^{2a}$, for $a=1,\dots 2d$. We require moreover that
\begin{equation}
[X^a,X^b]=0\; .\label{eq:comm}
\end{equation}
The model has a gauge invariance under $SU(N)$ transformations, where we act by
conjugation on all the matrices simultaneously $X^a\to U X^a U^{-1}$.
It is easy to see that the constraint (\ref{eq:comm}) is invariant under these transformations.

Now we want to solve a Gaussian matrix model quantum mechanics associated to these matrices.
We have two options here: we can consider a first order dynamics where $\bar Z^a$ and $Z^a$
are canonically conjugate variables (this is equivalent to stating that $X^{2a-1}$ and
$X^{2a}$ are canonically conjugate), or we can consider a second order dynamics where
we also include the time derivatives of the matrices (this doubles the number of matrices
effectively).  The first option is appropriate if we consider pure BPS states, since in that case $\Pi_Z =  i \bar{Z}$. However in our case we want to turn on some commutators between the fields, and thus go beyond the BPS spectrum. Therefore, we need to consider second order dynamics. The Hamiltonian will look as follows
\begin{equation}
H = \frac 12 \tr(\Pi_a^2) + \frac 12 \tr[(X^a)^2]\; .
\end{equation}

Because we have a $SU(N)$ action which leaves the model invariant, we can gauge  it,  and we can ask about the singlet sector of the matrix model. This is
the model we will concern ourselves with. We can now exploit the fact that the
matrices $X^a$ are Hermitian to use a $SU(N)$  transformation to diagonalize any one
of the them, let us say $X^1$. Because the matrices commute, they can be
diagonalized simultaneously, so if we diagonalize $X^1$, we diagonalize all others at
the same time.

This reduces the number of degrees of freedom to the eigenvalues of the matrices. Indeed,
for each diagonal component of the matrices $( X^a)^i_i$ we can associate a $2d$ vector of
eigenvalues
\begin{equation}
\vec x_i \simeq ( X^a)^i_i\; .
\end{equation}
In this form we have removed all of the infinitesimal gauge transformations on the $X$.
However, there are global transformations which permute the eigenvalues of the matrices
at the same time. These gauge transformations permute the vectors $\vec x_i$ into each other.
Because of this fact, wave functions have to be symmetric under the permutations of the
vectors $\vec x_i$.

The system can thus be interpreted as set of $N$ bosons on a space with $2d$ dimensions
(or a $2d$ dimensional phase space). If we treat the system classically, we can use a diagonal ansatz to find solutions of the
dynamical system. Under these assumptions we find $N$ free harmonic oscillators in $2d$
dimensions, which should be treated as $N$ identical particles (bosons) on a $2d$
dimensional harmonic oscillator.

Quantum mechanically, we can not do that immediately. This is because there are measure
factors that arise from the volume of the gauge orbit, and which affect the dynamics
of the system. This measure factor has been computed in \cite{Droplet}. It is given by
\begin{equation}
\mu^2 = \prod_{i<j} |\vec x_i-\vec x_j|^2 \; .
\end{equation}
and the reduced Hamiltonian is
\begin{equation}
H= \sum_i -\frac 1{2\mu^2} \nabla_i \mu^2 \nabla_i + \frac 12 |\vec x_i|^2 \;.
\end{equation}

We are now interested in studying the ground state wave function of the system and solving
the system in the thermodynamic limit. It turns out that
\begin{equation}
\psi_0 \sim \exp(-\sum \vec x_i^2/2)\;,
\end{equation}
is an eigenfunction of the above Hamiltonian. Since it is real and positive it is very likely
that it represents the ground state of the system. This will be orthogonal to other wave
functions of different energy by using the measure $\mu^2$. Namely, let $\tilde \psi$ be
another eigenstate of $H$ with different energy. Then
\begin{equation}
\int \prod_{i = 1}^N d^{2 d} x_i\,  \mu^2 \tilde \psi^* \psi =0\; .
\end{equation}

Moreover, vevs of observables will be evaluated with $\mu^2$, and this in general makes it hard
to do a calculation. It is convenient to perform a similarity transformation and absorb a
factor of $\mu$ into the wave functions, so that
\begin{equation}
\hat \psi = \mu \psi\;,
\end{equation}
and the measure factor associated to $\hat\psi$ is the usual $\prod d^{2 d}x_i$, which is $N$
copies of the measure for a single eigenvalue. Notice that $\mu$ is the square root of a
function which is symmetric in the exchange of all the vectors $\vec x_i$. So if the
particles are bosons with respect to the measure $\mu$, the particles given by the wave
function $\hat\psi$ are also bosons, with the usual measure for each boson. This regularity,
where we have $N$ identical copies of the measure factor of an individual boson, makes it
possible to treat the system thermodynamically, because we can place all bosons on the same
phase space and ask about the distributions of particles.

Now we want to study the large $N$ limit of the distribution of these bosons for the wave
function $\hat \psi$. If we square $\hat \psi$, we get a probability distribution on the
phase space of the $2N$ particles. This is given by
\begin{equation}
|\hat \psi_0^2| \sim \mu^2 \exp ( -\sum x_i^2 )
= \exp \left(-\sum \vec x_i^2 + 2\sum _{i<j} \log|\vec x_i-\vec x_j|\right)\;.
\end{equation}
The last term of the right hand side can be interpreted as partition function of a gas of
particles in an external quadratic confining potential,  $\exp(-\beta \tilde H)$, which has
logarithmic repulsion between the particles in $2d$ dimensions. In the thermodynamic limit
$N\to \infty$, we believe that the bosons will form some type of continuous distribution
density $\rho$ on the phase space of a single particle. The goal for us is to determine
the shape of $\rho$.

For $d=1$, this is a Coulomb gas in two dimensions, and the problem can be treated like a
plasma. The particles move to cancel the electric field locally, and they form a filled
disc of finite radius.  If the 2 dimensions are treated as a phase space, the system can be
related to a quantum hall droplet system of free fermions \cite{toyads}.

Now we will consider the case $d>1$. The probability distribution is given by
\begin{equation}
|\hat \psi_0^2| \sim \exp\left( - \int d^{2d} x \rho(x)  \vec x^2 + \int d^{2d} x
d^{2d} y \rho(x)\rho(y) \log|\vec x -\vec y| \right)\;.\label{eq:var}
\end{equation}
where $\rho$ is positive (the density of bosons) and  the total number of particles is  $N$.
This is imposed by the constraint  $\int \rho = N$. Now, we want to evaluate the distribution $\rho$ by a saddle
point approximation. The idea is  to treat the problem as a variational problem for
$\rho$ where we want to maximize the value of $|\hat \psi_0^2|$ which is our probability
density. We impose the condition of the number of particles as a constraint with a Lagrange
multiplier. We find that on the support of $\rho$
\begin{equation}
\vec x^2 + C = 2 \int d^{2d} y \rho_0(y) \log|\vec x - \vec y|\;.
\end{equation}
In general, one can show that for even numbers of dimensions the function
$\log| \vec x - \vec y|$ is proportional to the Green's function for the
operator $(\nabla^2)^d$, so that operating on both sides of the equation with this operator
one finds that for $d>1$, $\rho_0$ vanishes. This is incompatible with the
constraint that $\int \rho_0 = N$. This is what we find under the assumption that
$\rho_0$ is a smooth function.

What we find this way is that $\rho_0$ has singular support. Because of spherical symmetry,
one can make a simple ansatz for $\rho_0$ which has singular support. One takes a singular
spherically symmetric distribution at uniform radius $r_0$
\begin{equation}
\rho_0 = N \frac{\delta(|\vec x|- r_0)}{r_0^{2d-1} {\rm Vol} (S^{2d-1})}\;,
\label{rho0}
\end{equation}
which has been properly normalized. One sees this by transforming the integral
$\int d^{2d} x \rho_0(x) $ to spherical coordinates.

Now we substitute this ansatz into (\ref{eq:var}), and minimize with respect to $r_0$.
Since all particles end up at the same radius $r_0$, the term with $\int \rho_0 (x) \vec x^2$
is easy to evaluate. We find that this is equal to $N r_0^2$. The second term
is harder to evaluate. This requires integrating over relative angles. The term with the
logarithm is equal to $\log[ r_0 (1-\cos\theta)]$ for $\theta$ the relative angle between two
points on the sphere. This term can be written as follows
\begin{eqnarray}
\!\!\!\!\!\!
T_2(r_0)&=& N^2\! \int\! d^{2d}x d^{2d} y \rho(x)\rho(y) \log|\vec x - \vec y|
\\
&=&N^2\!\int\! \frac{d\Omega_{2d-1}d\Omega^\prime _{2d-1}}{\Vol(S^{2d-1})^2}
dr dr'{\delta(r-r_0)}
{\delta( r'- r_0)}\left[ \log(r_0)+\log(1-\cos\theta)\right] \;.\nonumber
\end{eqnarray}
Notice that in the above equation, only the first term of the sum will depend on $r_0$,
while the complicated angular integral will be in the second term of the sum.
Thus we find that $T_2(r_0)$ is equal to
\begin{equation}
T_2(r_0) = N^2\log(r_0) + N^2 c\;,
\end{equation}
where $c$ is a constant, independent of $r_0$. From here, we need to minimize the function
\begin{equation}
f(r_0) = N r_0^2  - N^2 \log(r_0) - N^2 c\;,
\end{equation}
from where we find that
\begin{equation}
r_0 = \sqrt {\frac N2}\;.
\label{rad}
\end{equation}
Notice that this result is independent of $d$. At first, this seems puzzling, but one can
argue that this is the correct result by calculating the force  particle $i$ exerts on
particle $j$ in the direction normal to the sphere.

\myfig{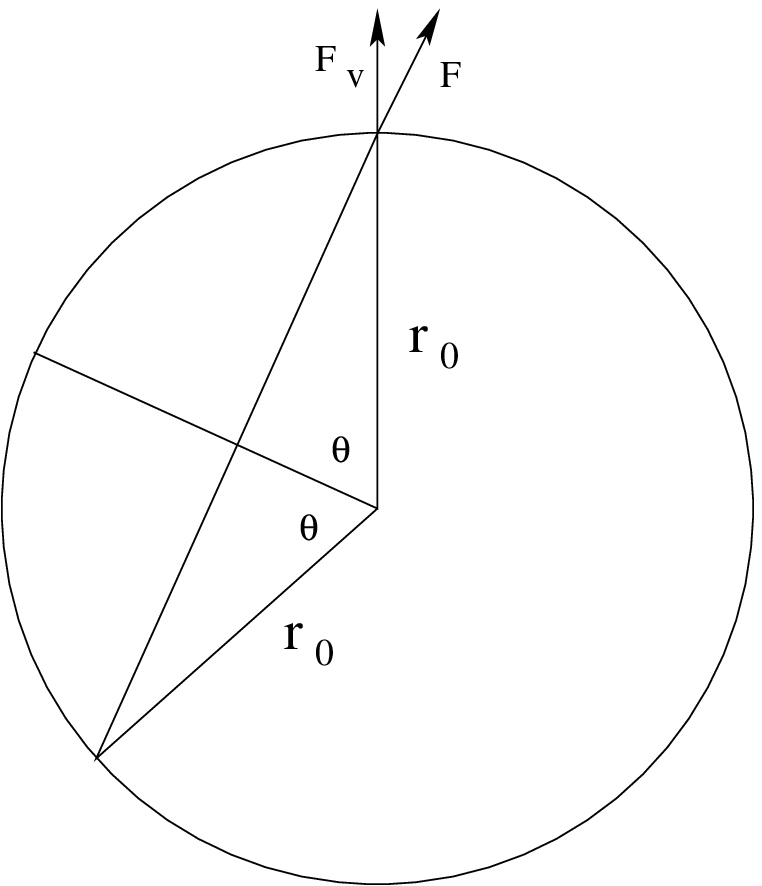}{6}{Diagram of forces for the statistical problem}

Looking at the figure \ref{fig:force.eps}, if the angle between the particles is $2\theta$,
then the distance between them is $l= 2r_0 \sin \theta$. The net force is then $2/l$ pointed
along the straight line joining particles $i$ and $j$. The normal to the sphere and this line meet
at an angle of $\pi/2-\theta$, and the force normal to the sphere from the particle at angle 
$2\theta$ (this force is pointing in the vertical direction in the figure) is then
\begin{equation}
F^{ij}_v = \frac {\cos(\pi/2-\theta)}{2 r_0 \sin\theta} = \frac 1{2r_0}\;,
\end{equation}
which is independent of the angle that particles $i,j$ subtend on the sphere, so long as
they are both located on the sphere. This is why the result does not depend on  $d$: the
angular distribution of particles (how many particles reside at angle $2\theta$) does not
matter to calculate the net force exerted on particle $j$.

The upshot of the above calculation is that the distribution of eigenvalues is a singular
distribution of particles. They form a thin shell of a sphere with radius $r_0$ independent
of $d>1$. The radius is exactly $\sqrt{N/2}$.

Now, we can use this radius to calculate various expectation values of various operators.
For example, let us take
\begin{equation}
\langle \frac 1N\tr X^{a_1} \dots X^{ a_{2n}} \rangle\;.
\end{equation}
This can also be calculated in the saddle point approximation used above. The $X$ variables
can be decomposed into spherical coordinates. In this form we obtain a relation
\begin{eqnarray}
\langle \tr X^{a_1} \dots X^{a_{2n}} \rangle
&\sim& \int \rho(x) x^{a_1} \dots x^{a_{2n}}\\
&=&
r_0^{2n} \int \frac{d \Omega_5}{\Vol (S^5) } \ \hat x^{a_1} \dots \hat x^{a_{2n}} \;,
\end{eqnarray}
where the $\hat x^{a_i}$ are unit vectors. Notice that this scales as $N^{n}$. This is the same
scaling that one obtains for one point functions in a single matrix model calculation. This
is suggestive that in the large $N$ limit vevs can be calculated systematically in a $1/N$
expansion for this ensemble. 
This would be interesting to understand. Notice also that these are symmetric
functions of the $a_i$. Similarly, in a Gaussian matrix model for the particle positions
$\vec x_i$, the radial integral would factor out, and we would obtain the same angular
integrals, which can therefore be evaluated using Feynman diagrams and gamma functions.

These techniques can also be used to represent excited wave functions. If we take a complex
combination of the fields (labeled as $Z^a$ above), one can take gauge invariant holormorphic
traces of the $Z^a$ and build states as follows
\begin{equation}
\hat\psi_{[n_1], [n_2],\dots,[n_k]}
= \tr(Z^{[n_1]})\tr(Z^{[n_2]})\dots \tr(Z^{[n_k]})\hat\psi_0\;,
\end{equation}
where the $n_k$ denote multi-indices.
These are conjectured to be eigenstates of the Hamiltonian of energy
\begin{equation}
\sum_j |n_j| \label{eq:energy}\;,
\end{equation}
above the ground state, which are moreover approximately orthogonal in the large $N$ limit \cite{Droplet}.
This follows from identifying these states with the corresponding graviton states in the
${\cal N}=4$ SYM theory. These give an approximate Fock space of oscillators, one for each
multi-index, on which one can take coherent states. These coherent states can be analyzed
using similar techniques as those used above, and they give wave-like shape deformations
of the five sphere, also with singular support in the embedding space $\BR^6$.

\section{A saddle point approximation to BMN state energies}
\label{spa}

Now we want to use the results of the last section to calculate energies of stringy modes in
the CFT. For this, we need an explanation of how the other modes of the SYM theory
decouple to obtain the matrix model of commuting matrices.
To do this we need to begin with the ${\cal N}=4 $ SYM theory compactified on a round $S^3$.
We obtain the following action for the scalars
\begin{equation}
S_{sc} = \int_{S^3} d\Omega_3\, dt \,  {\rm tr}
\left(\sum_{a=1}^ 6 \frac 12 (D_\mu\phi^a)^2-\frac 12 (\phi^{a})^2
-\sum_{a,b=1}^6\frac 14 g_{YM}^2 [\phi^a,\phi^b] [\phi^b, \phi^a] \right)\;.
\end{equation}
The mass term for the scalars is induced by the conformal coupling of the scalars to the
curvature of the $S^3$, which is chosen to have radius equal to one. This also sets the
scale for time derivatives. With this normalization, the volume of the $S^3$ is $2\pi^2$.

To study  BPS configurations, one needs to concentrate on the constant modes of the
$\phi^a$s, while keeping every other mode in the vacuum. This gives an effective reduction
to a gauged matrix quantum mechanical model of six Hermitian matrices. This model, after
rescaling the matrices to have a kinetic and quadratic potential term as in the last section,
is of the following form
\begin{equation}
S_{sc} = \int dt \,  {\rm tr}\left(\sum_{a=1}^ 6 \frac 12 (D_t X^a)^2-\frac 12 (X^{a})^2
-\sum_{a, b=1}^6\frac 1{8\pi^2}  g_{YM}^2 [X^a,X^b] [X^b,X^a] \right)\;.\label{eq:matmod}
\end{equation}
We will work with this dimensionally reduced model (slightly modified) in what follows.

The matrices at this point are not
required to commute. If we diagonalize $X^1$, and under the assumption that its
eigenvalues are of order $\sqrt N$ (as calculated in the previous section, and also as expected from
usual matrix integrals),
 we find that by putting vevs in the interaction term
coming from the commutators, the effective quadratic term for a generic matrix mode of any of the other $X^a$ is of order $g_{YM}^2 N$ larger
than the quadratic piece associated to  the free field result. This is true for matrix components that
do not commute with $X^1$. In the strong 't Hooft coupling limit this is a large number.
This means that the associated modes (which will be termed  off-diagonal) are very massive
and can be integrated out systematically. In a Born-Oppenheimer approximation these are treated as fast degrees of freedom, while we keep the information of the slow degrees of freedom exactly. This still leaves us with the configurations
where all matrices commute, they correspond to the matrix model described in the previous
section. We also cancel the zero point energy of the off-diagonal modes by hand, a fact that is expected by supersymmetry, although the exact cancellation mechanism could be quite involved in the details. This is a quantum modification of the matrix model, because we are not
keeping all the modes of the SYM theory, but we are keeping the induced quantum effects of 
the modes that we have integrated out. In the end, it is a prescription for a particular normal ordering of the off-diagonal modes, so we do not have to write a new effective action with these corrections made explicit, and we can work directly with \ref{eq:matmod}.

The idea that we will now pursue is that in the strong coupling limit of SYM, to look at
the lowest lying configurations (the ones associated to massless string states), we can
look at the reduced model described in the previous section and study the wave functions
in the reduced model. However, to include massive string states, we also need to consider
turning on these off-diagonal modes to a state which is not the ground state.

To do this systematically, we need to treat a particular off diagonal mode as a fast degree
of freedom attached to two sets of eigenvalues, which are slow degrees of freedom. The idea
is to treat the  off-diagonal modes perturbatively, and try to calculate the slow degrees of
freedom according to the commuting matrix  model description.  
In other words,  to a first approximation,  we ignore the backreaction that the off-diagonal modes produce on the geometry, and   treat them as modes of a free theory with Hamiltonian
\begin{equation}
\label{sbH}
H_{s b} =  \sum_{i \neq j} \frac{1}{2} (\Pi_a)_i^j (\Pi_a)_j^i  + \frac{1}{2} \omega_{i j}^2 (X^a)_i^j (X^a)_j^i \;,
\end{equation}
where the frequencies
\begin{equation}
\omega_{i j}^2 = 1 + \frac{g_{YM}^2}{2 \pi^2} |\vec x_i - \vec
x_j|^2\;,
\end{equation}
are evaluated using the classical distribution of eigenvalues of the matrix model discussed in the previous section. This is to be considered as a test of how well the commuting matrix model dominates the strong coupling dynamics of the SYM theory.

  In deriving (\ref{sbH}) we have used the fact that  the off-diagonal
modes between eigenvalues $i,j$ are to be treated orthogonal to the vector
$\vec x_i-\vec x_{j}$, because a component along that direction is obtained
by a gauge transformation on the commuting matrices. This is explained also in
\cite{Droplet}.


The above Hamiltonian can be regarded as a small perturbation in
the large $N$  't Hooft limit  as long  as $g^2_{YM} |\vec x_i -
\vec x_j|^2$  stays finite.  

Our approximation in what follows is that we will treat the off-diagonal modes as {\em free fields}, while we keep the information of the distribution of eigenvalues for the slow degrees of freedom exactly. For this approximation, the dimensional reduction of the matrix model is
a reasonable description of the system for the degrees of freedom we are considering, since we are ignoring interactions between off-diagonal modes. 

When we include the fact that the matrix model is gauged, we need to be careful about gauge invariance. This means that for each off-diagonal $X^j_k$ mode arriving at eigenvalue $j$, we need a second off-diagonal mode leaving it $X^l_j$. This means that the off-diagonal modes form a closed path between various points on the five-sphere. This is how we would like to
 think of a closed string state, where the off-diagonal modes can be labeled {\em string bits}. We will use this convention in what follows.

We now want to calculate the approximate energy of a state consisting of a single string bit joining two eigenvalues.  First,  we need a way to
identify which pairs of eigenvalues we are considering, and we also want to control the total
angular momentum of the string bit on the $S^5$, $J$.  To obtain states with the correct value of $J$, one also needs to
turn on the diagonal matrices to a state that is not the vacuum. This procedure can be
considered as a gravitational dressing of the state to impart it with momentum.
For large $J$, this will give us precisely the BMN limit \cite{BMN}.

We also need to make sure that we consider physical states of the gauge theory. Gauge
invariance forces us to turn on at least two such oscillators. One
from eigenvalues $i$ to  eigenvalues $j$, and the other from eigenvalue $j$ to eigenvalue $i$.
From our results in the previous sections, eigenvalues are going to be associated to positions
on the five sphere.

Let us now consider  a typical BMN-type operator
\begin{equation}
O_{k}  \sim \sum_{l = 0}^J \exp( i k l /J)\tr(Z^l Y Z^{J-l} X)\;.
\end{equation}
For $k\neq 0$ this is very similar to the operator
\begin{equation}
O_{k}  \sim \sum_{l = 0}^J \exp( i k l /J)\tr(Z^{l-1} [Y,Z] Z^{J-l-1} [X,Z] )\;.
\end{equation}
Using the operator state correspondence, we are supposed to relate the diagonal components
of $Z$ to the eigenvalues we found in the last section for the matrix $Z$. Let us call
these eigenvalues $z_i$. We do this because the state is almost BPS with energy approximately equal to $J$. The presence of the commutators means we are turning on
off-diagonal components of the fields $Y$ and $X$. These are to be treated as raising
operators in the quantum mechanical model (\ref{sbH}); call them $Y^{\dagger i}_j$ and $X^{\dagger k}_l$ for
the corresponding matrix modes.

We suggest that one treat the above operator as the following state in the reduced matrix model:
\begin{equation}
\ket{\psi_{k}} \sim \sum_{l = 0}^J \exp ( i k l /J) \sum_{j,j'}
z_j^l Y^{\dagger j}_{j'} z_{j'}^{J-l} X^{\dagger j'}_j \hat\psi_0 \vac_{od}\;,
\end{equation}
where in the above notation we have explicitly the wave function in the coordinate
basis for the diagonal components of the commuting part of the $X^a$ matrices, and where
we have the off-diagonal modes written as free oscillators acting on the off-diagonal
vacuum $\vac_{od}$.

Now we want to evaluate the energy of the above state.  We do this as follows:
\begin{equation}
 E \sim \frac{\bra{\psi_{k}} H^{total} \ket{\psi_{k}}}{\braket{\psi_{k}}{\psi_{k}}}\;.
\end{equation}
From the Hamiltonian (\ref{sbH}) we see that, after subtracting the ground state energy, each oscillator will carry an energy
\begin{equation}
E^{osc}_{jj'}= \sqrt{1+\frac{g_{YM}^2}{2\pi^2}  |\vec x_j-\vec
x_{j'}|^2}\;. \label{ode}
\end{equation}
Adding the energy of the diagonal piece by using  (\ref{eq:energy}) we get that the
total energy is given by
\begin{equation}
E^{total} = J + \langle E^{osc}\rangle\;, \end{equation}
 where we have to evaluate the average energy of the oscillator for the wave function we
 considered. This results into a multiple integral

\begin{equation}
\langle E^{osc}\rangle = \frac{\int \prod dx^i |\hat\psi_0|^2
\sum_{j,j'} |\sum_l \exp (ikl/J) z_j^l z_{j'}^{J-l}|^2 2
\sqrt{1+\frac{g_{YM}^2}{2\pi^2} |\vec x_j-\vec x_{j'}|^2}} {\int
\prod dx^i |\psi_0|^2  \sum_{j,j'} |\sum_l \exp (ikl/J) z_j^l
z_{j'}^{J-l}|^2}\;.\label{eq:energy2}
\end{equation}

In the above, we have done the contraction between raising and lowering operators of the off-diagonal modes as follows 
\begin{equation}
\langle a^{i}_{j} a^{\dagger k}_l \rangle = \delta^i_l\delta^k_l
\end{equation}
This is what makes the sums run over a single pair of eigenvalues in (\ref{eq:energy2}). The extra factor of $2$ in the above equation is due to the fact that we have two string bits between the same pair of eigenvalues in the problem.

Now we can evaluate this integral by a saddle point approximation, similar to what
we did in the previous section. This is done in two steps. First, the integral will be
dominated by configurations which maximize $|\psi_0|^2$, which suggest that $\vec x_j$
and $\vec x_{j'}$ should be located exactly on the sphere we found in the previous section.
Moreover, in the thermodynamic limit we can convert the sums $\sum_j$ into integrals
 $\int d^6x \rho_0(x)$, where $\rho_0(x)$ is distribution of eigenvalues (\ref{rho0}).
Thus we reduce the integral to an integral over relative angles associated to the positions
$j$ and $j'$ on the sphere, where we have identified locations on the sphere with individual
particles in our ensemble of the previous section,

\begin{equation}
\langle E^{osc}\rangle \simeq \frac{|\psi_0[\rho_0]|^2  \int
d\Omega_5 d{\Omega'}_5|\sum_l \exp (ikl/J) z^l {z'}^{J-l}|^2 2
\sqrt{1+\frac{g_{YM}^2}{2\pi^2} |\vec x-{{\vec x}'}|^2}}
{|\psi_0[\rho_0]|^2  \int d\Omega_5 d{\Omega'}_5|\sum_l \exp
(ikl/J) z^l {z'}^{J-l}|^2}\;.
\end{equation}
Using a description of the variable $z$ in spherical coordinates $z=r_0\cos\theta\exp(i\phi)$,  the square of the sum can be rewritten as
\begin{equation}
\left|\sum_{l = 0}^J \exp ( i k l /J) z^l {z'}^{J-l}\right|^2 =
r_0^{2J}\sum_{l=0}^{J}\sum_{l'=0}^{J}
(\cos\theta)^{l+l'}(\cos\theta')^{2J-l-l'} e^{i(l-l')\left(\frac{k}{J} +\phi' -\phi\right)}\;.
\label{sum}
\end{equation}

Now, we see that in the limit of $J$ large we can improve the
saddle point approximation due to the extra powers of $\cos\theta$
and $\cos\theta'$ in the numerator and denominator. First, the
angular integral  will be maximized when both of the $|z|$ take
their maximum value on the sphere. This implies that the integral
localizes, both on numerator and denominator on a particular
diameter of the sphere,  which is associated to null geodesics
associated to the $R$ charge carried by $Z$. For these points,
$|\vec x-{{\vec x}'}|^2$ reduces to $|z-z'|^2 = (2 r_0 \sin
{(\phi-\phi')/2})^2$,  where $\phi-\phi'$ is the angle along the
circle by which the two locations on the sphere associated to $j$
and $j'$ differ by. This is the difference on the arguments of the
two complex numbers $z$ and $z'$. Here we note that the directions associated to 
$X$ and $Y$ are orthogonal to the difference between these eigenvalues, so this is 
self-consistent 
with comments made previously.

Finally, the sum over relative phases in (\ref{sum}) can be approximated by a delta function
in the large $J$ limit. Both in numerator and denominator, the phase difference $\phi-\phi'$ 
is sharply peaked at $k /J$, because in the sum over phases the complex numbers align, 
while for other values of the angle, one sums over a lot of unit complex numbers pointing in all directions, which tend to cancel in the sum. From here, the effective energy of the oscillators is sharply peaked at
\begin{equation}
\langle E^{osc}\rangle = 2\sqrt{1+\frac { 2 g^2_{YM}  r_0^2}{\pi^2} \sin^2( k/2J)}\;,
\end{equation}
with the normalization of the wave function canceling between numerator and denominator. 
 Because the energy is sharply peaked the above state can be treated as an approximate 
 eigenstate of the full matrix model Hamiltonian.

The geometrical interpretation of this result is as follows. We have two eigenvalues on the sphere at a particular diameter, where the BPS null geodesics associated to the BMN limit for the corresponding configuration are located. The quasi-momentum on the BMN string, characterized by $k/2J$,  translates to the angle on the sphere between the two eigenvalues. The energy of the BMN impurities (the off-diagonal modes) is calculated by the Euclidean distance associated to the embedding of the five-sphere into a flat Euclidean 6-dimensional  geometry. This is shown in figure \ref{fig:geom.eps}.

\myfig{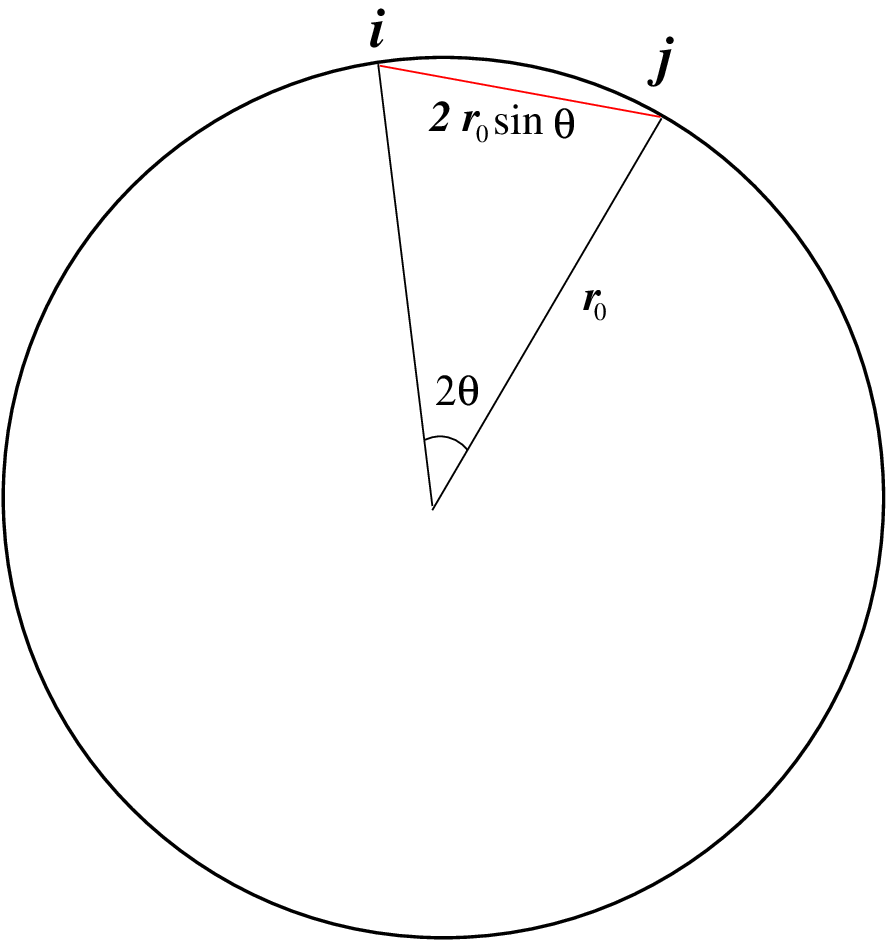}{4}{Geometry of the string bit between two eigenvalues, string bit shown in red. The angle between the eigenvalues is $\theta = k/2J$}

The energy of the string bit can be characterized in terms of its kinetic energy (coming from the free field ${\cal N}=4$ SYM result), plus the strong coupling mass term from the 
interaction with the eigenvalue condensate. This mass term is proportional to 
$|\vec x_i-\vec x_j|$  and not to the angle that the two eigenvalues on the sphere form. In this sense, the string bit leaves the sphere, but for small angle $\theta$ the effect is negligible.

Now we use the value of $r_0$ calculated in the previous section, and we find a result
which is equal to
\begin{equation}
\langle E^{osc}\rangle = 2\sqrt{1+\frac { g^2_{YM}  N}{\pi^2} \sin^2( k/2J)} \;.\label{eq:Etot}
\end{equation}
In the BMN  limit where $J$ is taken to be large,
we need to also take $k\sim 2\pi n$ with $n$ fixed. The above result reduces to
\begin{equation}
\langle E^{osc}\rangle = 2\sqrt{1+g^2_{YM}  N  \left(\frac{ n}{J}\right)^2}\;.
\end{equation}
which matches exactly the BMN limit to {\em all orders} in
perturbation theory \footnote{The normalization of $g_{YM}^2$
above is to be identified with $4\pi g$ in \cite{BMN}}. This
should be taken as evidence  that the approximations done above are reasonable for these states. 

Now we can try to compare our result (\ref{eq:Etot}) with some other conjectures
in the literature. Indeed, we find that the above analytic formula is exactly the
result of equation (2.23) in \cite{BDS} which is  an ansatz for  the energies of magnons  in a long range spin chain for an $SU(2)$ subsector.

We also have to take into account that we have made various approximations to obtain this result
and to discuss their range of validity. By inspection,  the geometry of the free field string bits is not what one would expect from
a semiclassical string embedded in the $S^5$, where the string bit energies should be measuring distances of curves tangent to the sphere. This can be traced back to the fact that
we treated the string bits as free fields. Indeed, this approximation seems to be valid for the BMN limit, but should not be valid for large angles on the sphere. One therefore expects that
having a fixed number of free string bits is going to be an incomplete picture of the string, and that one has to take seriously the possibility that the number of string bits is not a constant of motion. This seems to contradict the Bethe Ansatz conjectures of \cite{BDS}. Indeed, the number of impurities above the ground state is a conserved number in the Bethe Ansatz 
setup. However, we do obtain the energies associated to the quasi momenta of the excitations
exactly as described by their conjecture. This suggests that the Bethe Ansatz conjecture is 
incomplete, but at the same time it might be an intermediate step required to solve the full spectrum of strings on $AdS_5\times S^5$. This deserves to be investigated further. Indeed,  the Bethe-Ansatz 
predicts integrability, but the converse is not generally true.
 As a counter-example, 
one can consider the $c=1$ string, which is integrable, and where the number of incoming particles in the past is different than the number of outgoing particles in the far future.

\section{Discussion}
\label{dis}
In this paper we have shown how the eigenvalue distribution picture describing
BPS configurations of ${\cal N}=4$ SYM together with the non-BPS
off-diagonal modes can be used to faithfully reproduce some precise
geometrical calculations of strings moving in $AdS_5\times S^5$.

The main result of section \ref{mqm} is that the exact value
for the radius of the spherical distribution of eigenvalues was calculated. This radius
is $\sqrt{N/2}$ independently of the dimension, {\it i.e.} independently of the
number of matrices. This, that could seem unexpected a first sight, seems like a 
coincidence that is 
 necessary
so that the matrix model can capture the correct perturbative expansion of the field
theory, where to leading order the number of matrices/fields does not matter.

Then, we used a saddle point approximation for computing the
energy of BMN states to all orders in perturbation theory. We were
able to obtain an exact  agreement with the string 
theory result \cite{BMN} given the radius of the
$2d$-sphere (\ref{rad}) and the energy of the  off-diagonal modes
(\ref{ode}).  An interesting point about our results is the fact
that the wavefunction for the string bits becomes localized at a
particular ``classical" configuration in the large $J$ limit. This
matches the intuition that $J \rightarrow \infty$ is a
semiclassical limit \cite{GKP},  and that the spectrum of rapidly rotating
strings in $AdS_5\times S^5$ can be obtained by quantizing small
perturbations around classical string solutions \cite{Tseytlin,
Frolov}.

A
fundamental simplification in our computation came from treating the  string bits 
as modes of a free theory, which are geometrically understood as 
straight sticks connecting two points on $S^5$.
  This is a
reasonable approximation, as long as one is working in the strict
BMN limit. In that case, the length of string bits associated to
the off-diagonal modes shrinks to zero in the large $J$ limit and
our calculations show that it is correct to consider them as modes of a free
theory. On the contrary, to consider finite $J$ corrections (the
so-called near-BMN limit) one has to deal with string bits of
finite length. A long  string bit connecting two points on the sphhere, and several string bits joining
the same pair of eigenvalues via intermediate points are configurations with similar energy
in the strong coupling limit, as the distance between the eigenvalues dominates the
calculation of the energy.
 It is natural to expect that all these states will mix with
each other and the number of string bits will not be a good quantum
number \footnote{A simple model for variable number of string bits for open strings was found in \cite{BCV} in the study of giant graviton states}. In other words, string bits between eigenvalues separated by
finite distance in the 5-sphere cannot longer be treated as modes of
a free theory. Taking this into account is necessary if one is
interested in going beyond the BMN limit, either by considering a
finite $J$ or a large number of BMN excitations.
This result seems to be in conflict with the 
conjectures of Beisert, Dippel and Staudacher \cite{BDS}. We do seem to reproduce the
magnon energy for the Bethe ansatz they have devised, at the same time
we find that the number of impurities above the Bethe ground state is not conserved, and 
this appears to contradict the idea of building states using a generalized algebraic set of raising
operators as one would have on an algebraic Bethe ansatz approach. This seems to agree with
some observations of Minahan \cite{Mina} regarding the non-closure of the $SU(2)$ sector at strong coupling. It is not clear 
if our result is in conflict with the integrability of the string on $AdS_5\times S^5$ \cite{BPR,DNW},
 and it is possible that our results  coincide with the one loop Bethe ansatz \cite{MZ,BS} 
 when we take the small 't Hooft coupling limit. Indeed, our formulas seem to be analytic in the 't Hooft coupling for small coupling, a fact that suggests that one can match the perturbation theory 
 exactly if one works very hard.
 
 This will require understanding $1/J$ corrections very precisely.
 On top of what we have discussed above, 
another source of $1/J$ corrections in our calculation is the saddle point approximation
itself, because we obtained the expectation value of the energies by replacing
a complicated integral that depended on $J$ with an integral over a delta function distribution. 
The width of the correct distribution is related to $J$ and  
could produce corrections that need to be calculated. 
Calculating these corrections is  interesting also because they give rise to
disagreements between string theory and gauge theory computations
showing up at three loops in perturbation theory
\cite{Callan1},\cite{Callan2},\cite{Serban}.   It would be very
interesting to see
if the discrepancies between string and gauge theory beyond the BMN limit
can be overcome by extending the ideas presented in this paper.

Finally, we should point out that our discussion centered on the scalar s-waves of
${\cal N}= 4$ SYM. To understand better the structure of the superconformal multiplets 
at strong coupling, one 
would also want to do a similar analysis to the one we performed here 
with the spinors and the the gauge fields, plus the additional partial waves of all these fields
on the sphere.
This procedure might shed some light on how the different states mix, and it is probably a required
ingredient
to understand the possible integrability of the ${\cal N}=4$ SYM spin chain at strong coupling.

\section*{Acknowledgements}
D.B. would like to thank A. Tseytlin for some useful conversations.
D.B. work was supported in part by a DOE OJI award, under grant
DE-FG01-91ER40618. D.H.C. work
was supported in part by NSF under grant No. PHY99-07949 and by
Fundaci\'on Antorchas. S.E.V. work was supported by an NSF graduate
fellowship.

 \end{document}